\documentclass{article}

\usepackage[ruled,linesnumbered]{algorithm2e}

\usepackage{amsmath}
\usepackage{graphicx}
\usepackage{epstopdf}
\usepackage{PRIMEarxiv}
\usepackage{enumitem}
\usepackage[utf8]{inputenc} % allow utf-8 input
\usepackage[T1]{fontenc}    % use 8-bit T1 fonts
\usepackage{hyperref}       % hyperlinks
\usepackage{url}            % simple URL typesetting
\usepackage{booktabs}       % professional-quality tables
\usepackage{amsfonts}       % blackboard math symbols
\usepackage{nicefrac}       % compact symbols for 1/2, etc.
\usepackage{microtype}      % microtypography
\usepackage{lipsum}
\usepackage{fancyhdr}       % header
\usepackage{graphicx}       % graphics
\graphicspath{{media/}}     % organize your images and other figures under media/ folder
\usepackage{cite}

\title{HARL: A Novel Hierachical Adversary Reinforcement Learning for Automoumous Intersection Management
}

\author{
  Guanzhou Li \\
  Tsinghua University \\
  Beijing, China\\
  \texttt{ligz19@mails.tsinghua.edu.cn} \\
  \And
  Jianping Wu* \\
  Tsinghua University \\
  Beijing, China\\
  \texttt{jianpingwu@tsinghua.edu.cn} \\
  \And
  Yujing He \\
  Tsinghua University \\
  Beijing, China\\
  \texttt{hyj19@mails.tsinghua.edu.cn}\\
}

\begin{document}
\maketitle

\begin{abstract}
As an emerging technology, Connected Autonomous Vehicles (CAVs) are believed to have the ability to move through intersections in a faster and safer manner, through effective Vehicle-to-Everything (V2X) communication and global observation. Autonomous intersection management is a key path to efficient crossing at intersections, which reduces unnecessary slowdowns and stops through adaptive decision process of each CAV, enabling fuller utilization of the intersection space. Distributed reinforcement learning (DRL) offers a flexible, end-to-end model for AIM, adapting for many intersection scenarios. While DRL is prone to collisions as the actions of multiple sides in  the complicated interactions are sampled from a generic policy, restricting the application of DRL in realistic scenario. To address this, we propose a hierarchical RL framework where models at different levels vary in receptive scope, action step length, and feedback period of reward. The upper layer model accelerate CAVs to prevent them from being clashed, while the lower layer model adjust the trends from upper layer model to avoid the change of mobile state causing new conflicts. And the real action of CAV at each step is co-determined by the trends from both levels, forming a real-time balance in the adversarial process. The proposed model is proven effective in the experiment undertaken in a complicated intersection with 4 branches and 4 lanes each branch, and show better performance compared with baselines.

\end{abstract}

% keywords can be removed
\keywords{hierarchical learning, hierarchical adversarial learning, HARL, Reinforcement learning, Autonomous intersection management, Connected Autonomous Vehicle}

\section{Introduction}
In the urban road net, intersections are the major locations for traffic delay and accidents \cite{chen2021mixed}, where traffic flows from multiple directions converge and intertwine to form many conflicting points. Moreover, various traffic participants interact in this scenario in a complex way, which also makes it an accident-prone scene. Federal Highway Administration declares 40 percent of crashes were related to intersections. The advent of Connected Autonomous Vehicles (CAVs) provides a novel perspective for a safe and efficient passage at intersections. Real-time communication of CAVs offers new way to organize traffic flows at intersection, giving birth to the concept of Autonomous Intersection Management (AIM). Designed as self-organized system, AIM discusses the optimal crossing schedule for all vehicles at unsignaled intersections, which will reduce rule-mandated stops or decelerations in predefined traffic signal control. In terms of communication protocols and scheduling algorithms, methods of AIM can be categorized into centralized and distributed way. In the former, the intersection manager (IM) acquires status of each vehicle mainly through vehicle-to-infrastructure (V2I) communication and determines crossing order of vehicles globally, while centralized IM is unnecessary or less important in the latter way. Instead, CAVs talk to each other through vehicle-to-vehicle (V2V) protocol and adjust their speeds or trajectories according to self-decision algorithm or form a consensus-based crossing schedule. The pros and cons of these two methods is regarded as a trade-off between complexity and performance generally \cite{otto2009distributed}. Centralized AIM benefits from global observations and considerations thus it is easier to design a well-performed schedule, but it causes higher computational load, and lacks of robustness due to its dependencies upon completeness and accuracy of information. For example, clock synchronicity is required in the process of collecting messages from CAVs and making decisions otherwise collisions may occur due to differences in built-in clocks from vehicle to vehicle. In contrast, distributed AIM controls each CAV in a relatively independent way and reduces the need for synergy of movements, making it more robust. Nevertheless, there are probabilities of falling into sub-optimum restricted to the locality of reception and decision in distributed way. Hence, a delicate design is required to make algorithms work in complicated interactive scenarios.

Reinforcement learning (RL) has achieved success in many domains from robotics to games. However, it is not until recent that RL was introduced into AIM to coordinate CAVs \cite{wu2019dcl}. Learning-based AIM can produce scheduling plan faster compared to optimization-based way and is more flexible and well-performed compared to rule-based way. 
As the dimension of the action space varies with the number of vehicles in the crossing schedule in the centralized RL,making the output of RL hard to design, RL in AIM mostly adopts a distributed way. The agent in distributed RL obtains state from observation of each vehicle and navigate it through the intersection with control instruction. However, it is challenging to generate a generic strategy for all vehicles which interact with each other in complicated ways in distributed RL. The actions from the sample of a generic policy, taken by multiple sides of conflicts is prone to unavoidable collisions. DCL-AIM attempted to address the dilemma of vehicles with the same strategy by divide the state of vehicle into independent state and coordinated state \cite{wu2019dcl}. Li et al. formulated the interactions between vehicles as leader-follower gaming process \cite{li2020game}. The adv.RAIM encoded the states of surrounding vehicles into a representative feature and fed it to the agent to make decision better \cite{guillen2022multi}.

This paper proposes a novel perspective to regard this problem by applying a hierarchical adversarial reinforcement learning, agents in upper level devote to pull the vehicle out of being collided and agents in lower level make sure the change of mobile state will not cause new collisions. They observe environment from different receptive scope and make decision according to comprehensive consideration of reward feedback in different time periods in the future. In this driving way, the vehicles can adjust their speed more smoothly and perform better in collision-avoidance. Specifically, two groups of hierarchical structures are availed for preparation to find the proper time to pass before crossing the intersection and fine tune of speed to avoid collisions when crossing the intersection, respectively. A hierarchical adversarial RL framewokr with a Soft-Actor-Critic (SAC) model in upper level and one in lower level is involved for each group. The agents in different level acquire state in different receptive scopes and perspectives, and they accomplish actions and receive corresponding reward in different ranges of period. To reach a balance between collision-free and passing efficiency better, various reward mechanisms are configured for SAC in different levels. This will significantly improve the crossing safety as the reward of collisions is rarely acquired by agents in an effective scheduling. Besides, a shared expected reservation table provide message exchange among various CAVs thus they can interact with each other better before entering the intersection. The main contributions of this paper include:

\begin{enumerate}
    \item A novel hierarchical adversarial reinforcement learning is brought in to schedule vehicles in AIM, with multiple receptive scopes and reward-acquired step size.
    \item A expected reservation table is availed for interaction between multi agents before entering crossing zone, to search for the optimum passing time for each CAV, which significantly reduces the number of conflicts at intersection.
    \item We measure the performance of proposed model with several baselines in various metrics in a complex intersections.
\end{enumerate}

The rest of the paper is organized as follows. Chapter II reviews relevant literature in AIM problem, Chapter III illustrates details of HARL. Chapter IV describes the experiments and compares HARL with baselines. Conclusions and discussions are presented in Chapter V.

\section{Literature Review}
\subsection{Traffic Light and AIM}
Until nowadays, traffic lights still play an indispensable role in conducting traffic at intersections and numerous researches have been carried out around it. Despite long-term development,  the plan of traffic lights in many cities still deploy fix-time strategy around the world, and Webster signal setting is one of the most well-known approach \cite{webster1958traffic}. Apart from pre-defined timetable, various adaptive timing strategies like vehicle-actuated \cite{tyack1938street}, pressure-based methods have been launched \cite{gregoire2014back, boukerche2021novel}. The longest-queue-first policy, as one of the pressure-based methods, which always let the vehicles in the direction with longest queue pass, was proven efficient and robust \cite{wunderlich2008novel}.

The Autonomous Intersection Management (AIM), proposed by Drenser et al.\cite{dresner2004multiagent}, is the "traffic light" designed for emerging traffic participants with negotiating ability and lower reaction delay -- Connected Autonomous Vehicles (CAVs). Instead of observing the color of traffic light, CAVs gain right of crossing by requesting or being assigned \cite{khayatian2020survey}. In the request mode, CAVs send the messages about their expected speed or crossing time to the intersection manager (IM) then the IM collects all messages, and approves or rejects each request. While in the assign-based method, the IM actively assign space-time slots to vehicles for their passages. How to negotiate in a comprehensive way has been studied by \cite{bashiri2017platoon,aoki2018dynamic}. Complete and valid instruction set can help CAVs to crossing the intersection more safely and efficiently.Besides, ad-hoc technique overcomes limited communication coverage and shows significant improvement in driving safety \cite{joerer2012crash}.

The hybrid intersection refers to the intersection management for both CAVs and Human-driving vehicles (HVs)\cite{khayatian2020survey}. Despite the rapid increase in the market share of intelligent vehicles recently, HVs will remain a non-negligible traffic participant for a least the next few decades. Reports indicate that the market penetration rate of CAVs will not exceed 90\% by 2045 \cite{hernandez2015v}. Thus the scheduling problem in the hybrid intersection has attracted intensive studies. As one of the initial attempts, FCFS-Light associated the principle of First-Come-First-Serve with traffic light \cite{dresner2007sharing}. H-AIM filled the gap of FCFS-Light in the low CAV penetration condition by adjusting the rule of no-reservation green lane to no-reservation active green light when permitting or denying the request of right of crossing from CAV in red \cite{sharon2017protocol}. Besides, there are also fuzzy-logic control and reinforcement learning leveraged to address traffic scheduling problems in the hybrid intersection \cite{onieva2015multi,quang2020proximal}. As an adaptive algorithm, distributed RL is naturally suitable for scheduling in the hybrid intersection. The CAVs will dynamically adjust their strategies according to their own observations and received messages to avoid collide with HVs, removing the requirements to control all vehicles at the intersection in some rule-based algorithms.

\subsection{Scheduling Policy}
The AIM essentially includes two branches -- one branch is to re-order the sequence that vehicles entering or leaving intersection, and the other is to search trajectories along which vehicles can cross interweaving areas safely. We focus on the former branch in this paper. The objectives of scheduling consists of safety, efficiency, fairness, environment and comfort, among which the safety comes first and no conflict between vehicles needs to be guaranteed under any circumstances. Under this principle, the scheduling methods are categorized into First-In-First-Out (FIFO), heuristic, and optimization-based \cite{gholamhosseinian2022comprehensive}. FIFO, as the name implies, the earlier approaching vehicle has higher priority to leave the intersection. optimization-based method formulates the AIM as conditional optimizing process and seeks for the global optimum, while heuristic approach do the optimization in a simpler way by swapping or remaining the passing order between adjacent vehicles per time. Among these approaches, FIFO is the earliest studied and shows the most fairness, heuristic approach is relatively easy to operate, and optimization-based way is the most complicated but the most effective and gains the least delay so it is widely studied from various viewpoints. Further, we summarize the scheduling method into six kinds: operational planning, tree search algorithm, co-utility maximization, game theory and bidding system, learning-based approach, and others.

Specifically, in the first kind, some methods in linear optimization like Big M method, Brand-and-Bound, and mixed-integer linear programming (MILP) are introduced to find the solution \cite{jin2012multi,yang2016isolated,fayazi2018mixed}. Further, the MILP is extended to non-linear form as mixed-integer non-linear programming for collision-free plan \cite{mirheli2019consensus}. Besides, other approaches in this kind involve Lagrange function, Euler-Lagrange equation and inexact Newton method \cite{belkhouche2018collaboration, malikopoulos2019closed, wang2020cooperative, jiang2017distributed}. This group of methods formulates the scheduling problem at intersection as conditional mathematical optimization problem and devotes to solve it. The accuracy of modeling the real situation and the accuracy and time consumption of the solution are the key factors that affect the performance of such methods.

In the tree algorithms, each leaf corresponds to an efficient schedule and each vehicle search the passable time slots by traversing the tree. Since Li et al. put forward a tree algorithm in the AIM problem \cite{li2006cooperative}, many different way to build a tree to formulate the scheduling problem has been proposed, including adaptive belief tree under uncertain environment, red-black tree, and Monte-Carlo tree search \cite{hubmann2017decision,choi2019reservation,xu2019cooperative,mirheli2018development}. The tree structure makes the process of decision more tractable and help the system to reach the optimum step by step, but the tree will become deep and large when the scenario is complicated, making it challenging and costly to remain, update, and traverse the whole tree.

The co-utility maximization regards the AIM from a system perspective and chases the maximization of social benefits. When approaching intersection, each CAV is required to report its utility to IM, and IM assigns road priority according to the collective utility, or namely collective welfare. The objective function of collective welfare can be optimized through both heuristic and optimization-based way \cite{buckman2019sharing,sayin2018information}. The concept of collective welfare is interesting but the results will highly rely on the definition of utility function, and the effect on one vehicle from another in the interactions is hardly reflected from total utility alone.

If say co-utility is a centralized concept, the game theory considers the AIM from a distributed viewpoint. It describes the interaction among multiple game players and is fit to characterize the decision-making process between CAVs when they are in conflict with each other. Chicken games was first introduced to determine acceleration or deceleration between two conflicting CAVs by \cite{elhenawy2015intersection}, and then Rahmati availed game-theory in the left-turn maneuvers of CAVs in the unprotected environment \cite{rahmati2017towards}. Further, Wei et al. developed multi-layer games involving platoon formation game and collision-avoidance game \cite{wei2018intersection}. Apart from game theory, auction mechanism is also a competitive strategy, the CAV bid constantly in auction until defeat its competitor and acquire right of way \cite{vasirani2012market}. The bidding system was further developed to auction between groups, in the WIN-FIT model, CAVs dynamically form groups spontaneously and devote to beat other group, and the winner is permitted to pass through the intersection preferentially then other fit into their gap safely \cite{chen2015win}. 

The learning-based approach is an emerging technology. As mentioned in Chapter I, RL has recently emerged as the mainstream of learning-based approach. To the best of the author's knowledge, most of them use distributed RL. Distributed RL was first introduced in AIM by \cite{wu2019dcl} by availing joint Q-values at coordinated states allowing for sparsely coupling between agents. Game theory was utilized to depict the coordination between agents in \cite{li2020game}. However, the aforementioned works mainly considers the pairwise interaction between agents and pairwise cooperation faces challenges deploying in frequently multi-party interactive scenarios. For the multi-party interaction, \cite{guillen2022multi} proposed adv.RAIM exploiting encoder neural networks to embed high dimensional states to feature space for each vehicle to learn and react.

Finally, Some other studies are focused on fuzzy logic \cite{onieva2015multi,abdelhameed2014hybrid,abdelhameed2014development}, particle swarm optimization \cite{guney2020scheduling,saust2012energy}, etc.

\section{Methodology}
\subsection{Problem Statement}
\begin{figure}[t]
    \centering
    \includegraphics[width=0.8\textwidth]{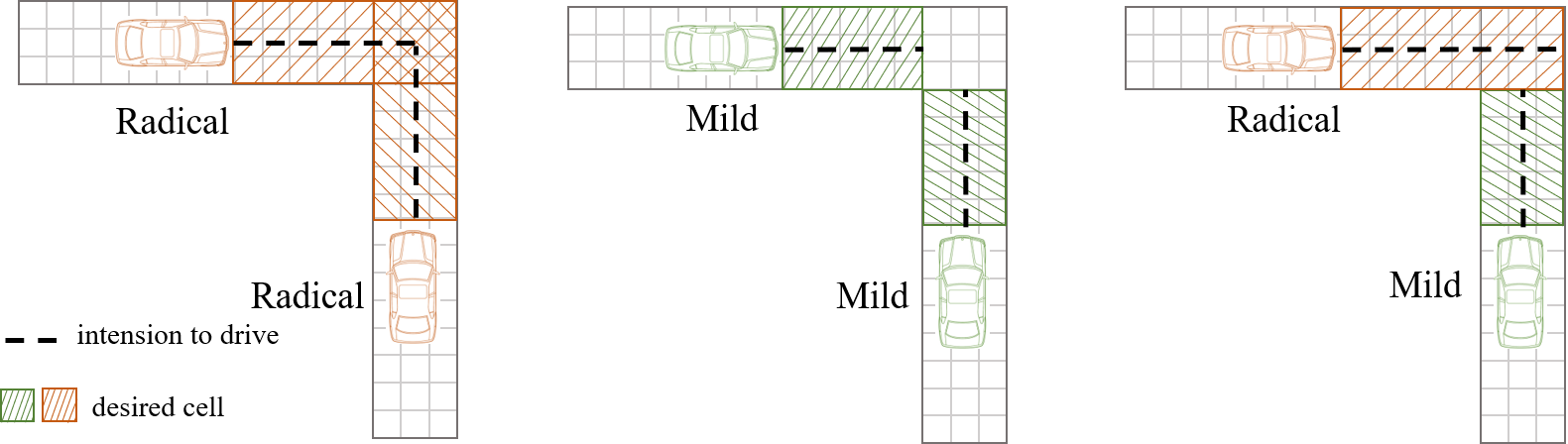}
    \caption{Interactions between CAVs as different agents}
    \label{fig:fig1}
\end{figure}

Comparing to widely studied rule-based scheduling optimization, RL encounters difficulties and is relatively rare deployed in AIM as the interactions between vehicles cause fluctuations and divergence of results. As the number of CAVs involved in AIM varies in different scenarios, the size of action space of centralized RL is uncertain and hard to design, which makes distributed RL more suitable for the AIM than centralized RL. For the same reason of quantitative uncertainty in CAVs, distributed RL requires to sample action from a generic policy with sharing parameters. Nevertheless, the complicated interactions at intersection and local decisions of distributed RL makes it difficult to avoid conflicts between CAVs when crossing intersection. Besides, action selection from the same policy is prone to collision when two CAVs in similar states meet, as shown in the figure \ref{fig:fig1}. Therefore, the learned strategy requires the ability to adjust actions adaptively to the opponent's movements to avoid collisions, and the different parties in the encounter need to respond differently so as not to get caught in a dilemma cycle. There are studies attempting to address these challenges by introducing game theory or cooperation mechanism into RL framework. But their decisions and actions focuses mainly on the interaction between pairs of CAVs, efficiency and safety cannot necessarily be guaranteed in complicated situations where multiple CAVs interact simultaneously. We construct a multi-agent framework to integrate trends arising from different intentions and these trends co-determine the final action of vehicle at each step.

\subsection{Soft Actor Critic}
In the RL framework, agent interacts with environment by sampling action from strategy $\pi(a_t |s_t)$ at step $t$, and acquires corresponding reward $r_t$ from environment then transfer to next state $s_{t+1}$. The process keeps shaping agent's principle of actions until it finds an optimal policy, and it can be formulated as a Partially Observable Markov Decision Process (POMDP) in AIM defined by five-variable tuple $(S,A,P,R,\gamma)$, where state $S$ and transmission probability $P$ are given by environment, $A$ denotes action space of agent, and $R$ gives reward or punishment to each action, the $\gamma$ discount future reward to prevent infinite total reward.

Deep Reinforcement Learning (deep RL) utilizes deep learning to estimate utility of states and policy of actions in RL, which can be categorized into value-based RL represented by Deep Q learning and policy-based RL represented by actor-critic. Value-based deep RL approximates the value function of each state and the agent selects the action with maximum value, while policy-based RL attempts to figure out the optimum distribution of actions under certain state, which is more flexible to deal with states unseen by agent before and uneasy to get stuck in a local optimal trajectory. The countless interactive patterns cause nearly infinite states in AIM problem. Some states differ little in value but the results of the same action will vary greatly like from thrilling passage to collisions, making probabilistic action model in policy-based RL a better choice than deterministic action from value-based RL here. As an advanced off-policy actor-critic algorithm, Soft Actor Critic is capable of exploring more actions with entropy maximization term to encourage exploration in early stage. In SAC, the policy updated in each step is proven better than the
previous step\cite{haarnoja2018soft} hence the property enables the agent improve their ability stably.

The state of SAC agent comes from the observations of each CAV, including their own and surrounding vehicles' movement characteristics, like position, speed, angle, lane, and action given by other agent. The state at step $t$ is written as $s^{(j)}_{i,t}$, where $j$ is the marker of a CAV in the scheduling list of CAVs, and $i$ denotes the CAV is considered in the perspective of $i$-$th$ agent. For the convenience of expression, the aforementioned CAV considered by $i$-$th$ agent is marked as $H^{(j)}_i$. Then the action of $H^{(j)}_i$ is given by a mapping from $s^{(j)}_i$, representing the expected speed change in the next action step of $i$-$th$ agent, restricted within action space $[a_{i,min},a_{i,max}]$.
\begin{equation}
    a_{i,t}^{(j)}=
    \begin{cases}
    random(a_{min},a_{max}), &\text{if} \ \ t \leq \text{start time of action}\\
    f_{\phi_i} (\epsilon_{i,t}^{(j)};s_{i,t}^{(j)}) &\text{if} \ \ t > \text{start time of action}
    \end{cases}
\end{equation}
where $\phi_i$ is the parameters of actor networks of the $i$-$th$ SAC agent and $\epsilon_{i,t}^{(j)}$ denotes randomly generated Gaussian noise, $s_{i,t}^{(j)}$ is the current state. After taking action $a_{i,t}^{(j)}$, $H_{i,t}^{(j)}$ will obtain reward $r_{i,t}^{(j)}$ and the environment will come to next state $s_{i,t+\Delta t}^{(j)}$, with $\Delta t$ the action step size of $i$-$th$ agent. The experience of training is given by a five element tuple: $(s_{i, t}^{(j)}, a_{i, t}^{(j)}, r^{(j)}_{i,t}, s_{i, t}^{(j)},\delta_{i,t}^{(j)})$, where $\delta_{i,t}^{(j)}$ is a signal indicating whether the $i$-$th$ agent has finish its task for the CAV at step $t$. Each agent has its own memory palace $D_i$ updated with the experience of all CAVs controlled by the $i$-$th$ agent $H_i^{(1)},H_i^{(2)},…,H_i^{(k)}$ in the following way: 

\begin{equation}
    D_{i}=D_{i} \cup\left(\cup_{j=1}^{k}\left\{\left(s_{i, t}^{(j)}, a_{i, t}^{(j)}, r^{(j)}_{i,t}, s_{i, t}^{(j)},\delta_{i,t}^{(j)} \right)\right\}\right)
\end{equation}

The training process of each SAC agent involves updates of a Value network, a Q network, and a Policy network. The Value network gives the value estimations of states and updated with objective function:
\begin{equation}
    J_{V}\left(\psi_{i}\right)=\mathbb{E}_{\mathbf{s}_{t} \sim \mathcal{D}_{i}}\left[\frac{1}{2}\left(V_{\psi_{i}}\left(\mathbf{s}_{t}\right)-\mathbb{E}_{\mathbf{a}_{t} \sim \pi_{\phi_{i}}}\left[Q_{\theta_{i}}\left(\mathbf{s}_{t}, \mathbf{a}_{t}\right)-\log \pi_{\phi_{i}}\left(\mathbf{a}_{t} \mid \mathbf{s}_{t}\right)\right]\right)^{2}\right]
\end{equation}
Where $V_{\psi_i} (s_t), Q_{\theta_i} (s_t,a_t), \pi_{\phi_i} (a_t|s_t)$are the values yielded from Value Network, Q Network and Policy Network in $i$-$th$ agent, and $\psi_i,\theta_i,\phi_i$ denote the hyper-parameters of those networks, respectively. $(s_t,a_t)$ is a state-action pair sampled from memory $D_i$.
Subsequently, the objective function of the Q network for $i$-$th$ agent is expressed as:
\begin{equation}
    J_{Q}\left(\theta_{i}\right)=\mathbb{E}_{\left(s_{t}, \boldsymbol{a}_{t}\right) \sim \mathcal{D}_{i}}\left[\frac{1}{2}\left(Q_{\theta_{i}}\left(\boldsymbol{s}_{t}, \boldsymbol{a}_{t}\right)-\left(r\left(\boldsymbol{s}_{t}, \boldsymbol{a}_{t}\right)
    +\gamma \mathbb{E}_{s_{t+1} \sim p}\left[V_{\bar{\psi}_{i}}\left(\boldsymbol{s}_{t+1}\right)\right]\right)\right)^{2}\right]
\end{equation}
Where \(\hat{\psi}_i\) is an exponentially moving average of value network weights updated with \(\hat{\psi}_i\leftarrow \tau\psi_i+(1-\tau)\hat{\psi}_i\) to stabilize the training. And the objective function of policy network for $i$-$th$ agent is shown as:
\begin{equation}
    J_{\pi}\left(\phi_{i}\right)=\mathbb{E}_{s_{t} \sim \mathcal{D}_{i}, \epsilon_{t} \sim \mathcal{N}}\left[\log \pi_{\phi_{i}}\left(f_{\phi_{i}}\left(\epsilon_{t} ; \boldsymbol{s}_{t}\right) \mid \boldsymbol{s}_{t}\right) -Q_{\theta_{i}}\left(\boldsymbol{s}_{t}, f_{\phi}\left(\epsilon_{t} ; \boldsymbol{s}_{t}\right)\right)\right]
\end{equation}

Following the work in \cite{haarnoja2018soft}, automated tuning of temperature parameter are utilized during training.

\subsection{Hierarchical adversarial mechanism}

\begin{figure}[t]
    \centering
    \includegraphics[width=0.8\textwidth]{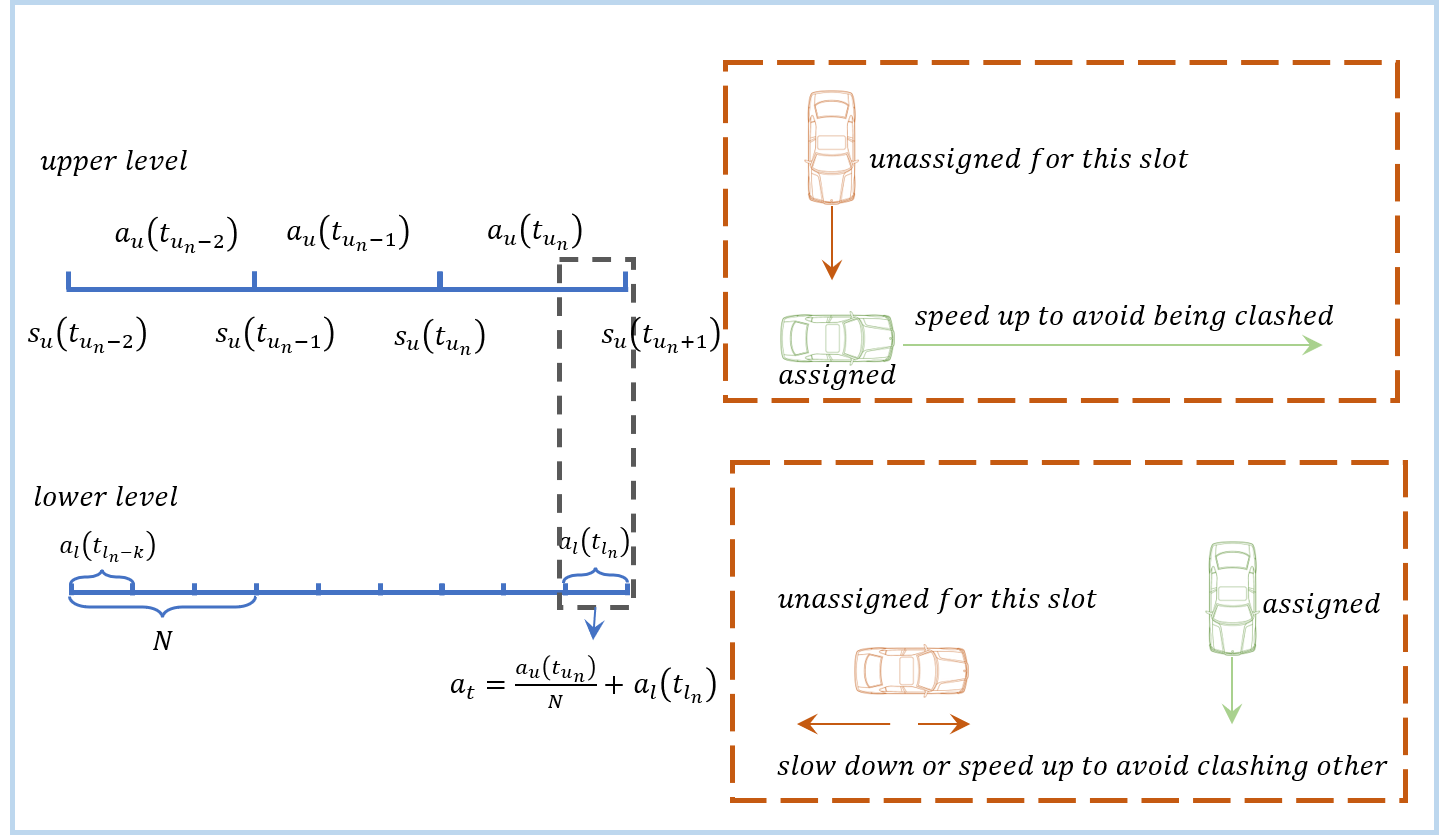}
    \caption{hierarchical adversarial mechanism}
    \label{fig:fig2}
\end{figure}

To avoid conflicts and collisions between vehicles from complex interactions at intersections, a hierarchical framework is introduced to train deep RL with the agents at upper level attempt to prevent being clashed by others currently or in the near future and the agents at lower level guarantee the behavior caused by the upper level will not collide or conflict with other vehicles. Intuitively speaking, the trends of actions from upper level push CAVs forward and the actions from lower level pull CAVs back, a proper action at each step will be produced from their balance in the adversarial process \ref{fig:fig2}. 

\begin{figure}[t]
    \centering
    \includegraphics[width=0.8\textwidth]{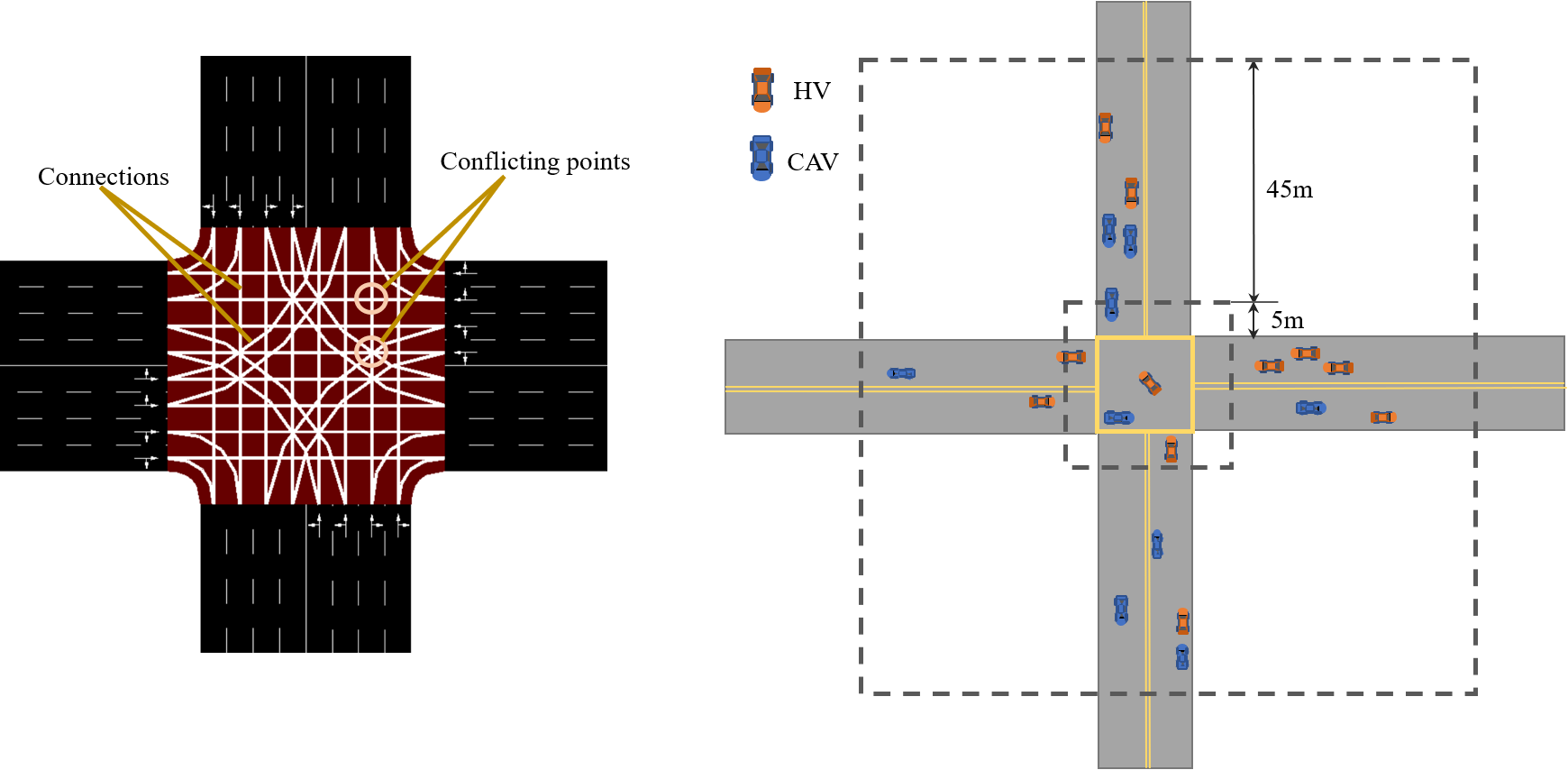}
    \caption{Intersection diagram}
    \label{fig:fig3}
\end{figure}

To start with, we define all vehicles wishing to cross the intersection with $\hat{H} = {[H^{(1)},H'^{(2)},\cdots,H^{(j)},\cdots,H'^{(j')},\cdots]}$, where $H^{(j)}$ denotes a CAV in the list with $j$ its index, and $H'^{(j')}$ represents the human-driven vehicle. Each vehicle $H^{(j)}$ drives in a connection expressed by $C = $ (Entrance Lane, Exit Lane), which passes a series of conflicting points $Q_c = {[q_{c1},q_{c2},...,q_{cm},...]}$ as shown in figure \ref{fig:fig3}. Further, a expected reservation time table $M \in {\{0,1\}}^{N_q \times T_r}$ is availed to record the expected crossing time slots of vehicles, where $N_q$ is the length of $Q_c$ and $T_r$ is the length of future time-step to be considered for $M$. When specified slots are occupied, the corresponding elements in $M$ are set to $1$ otherwise $0$.
 The allocation of time slots of each vehicle in $M$ is followed the algorithm \ref{alg:alg1}, and the set of conflicting vehicles in this step, noted as $Lc$, will be updated in this process. For the convenience of expression, we use $H''^{(j)}$ to refer to a vehicle which can be either a CAV or a HV.
\begin{algorithm}[h]
    \caption{update $M$ and $Lc$ with the vehicle $H''^{(j)}$}
    \label{alg:alg1}
    \SetKwInOut{Input}{Inputs}
    \SetKwInOut{Output}{Outputs}
    \SetKwInOut{Definition}{Definition}
    \SetKwInOut{Initialize}{Initialize}
    \Definition{$D^{(j)} := [d_{c1},d_{c2},...,d_{cm},...]$ with $d_{cm}$ the distance from current position of $H''^{(j)}$\\
                to the conflicting points $q_{cm}$\\
                $v^{(j)} := $ the current velocity of $H''^{(j)}$\\
                $l^{(j)} := $ the length of vehicle}
    \Input{$M$,$Lc$,$D^{(j)}$,$v^{(j)}$,$l^{(j)}$}
    \Output{$M$,$Lc$}
    \Initialize{$\delta_{c}=False,tmp=\{\}$}
    \For{$k=1,2,...,m,...$}{
        $ind_{start} = d_{ck}/v^{(j)}$\\
        $ind_{end} = (d_{ck} + l^{(j)})/v^{(j)}$\\
        \If {($ind_{start}\geq0$ and $ind_{start}\leq T_r$) or ($ind_{end}\geq0$ and $ind_{end}\leq T_r$)}{
            $ind_{start} = max(min(ind_{start},T_r-1),0)$\\
            $ind_{end} = max(min(ind_{end},T_r-1),0)$\\
            \If {$sum(M(ck,ind_{start}:ind_{end})) > 0$}{
                $\delta_{c}=True$\\
                $H''^{(j')} \leftarrow $ get the vehicle occupying slot\\ $M(ck,ind_{start}:ind_{end})$\\
                $tmp = tmp.union(\{H''^{(j)},H''^{(j')}\}$
            }
            $M(ck,ind_{start}:ind_{end}) = 1$
        }
    }
    \If{$\delta_{c}$ is True}{
        $Lc = Lc.union(tmp)$
    }
    \textbf{return} $M$, $Lc$
\end{algorithm}

\begin{figure}[t]
    \centering
    \includegraphics[width=\textwidth]{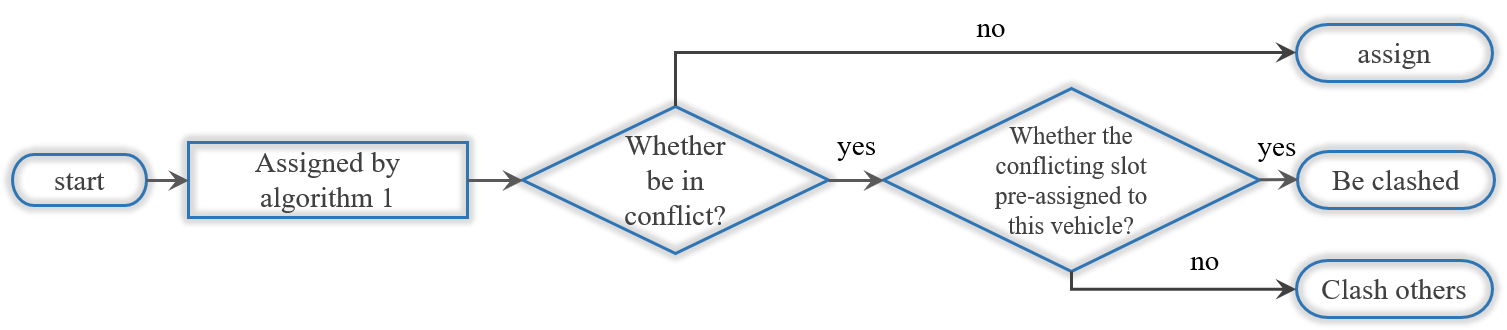}
    \caption{assignment mechanism}
    \label{fig:fig4}
\end{figure}

\begin{figure}[t]
    \centering
    \includegraphics[width=\textwidth]{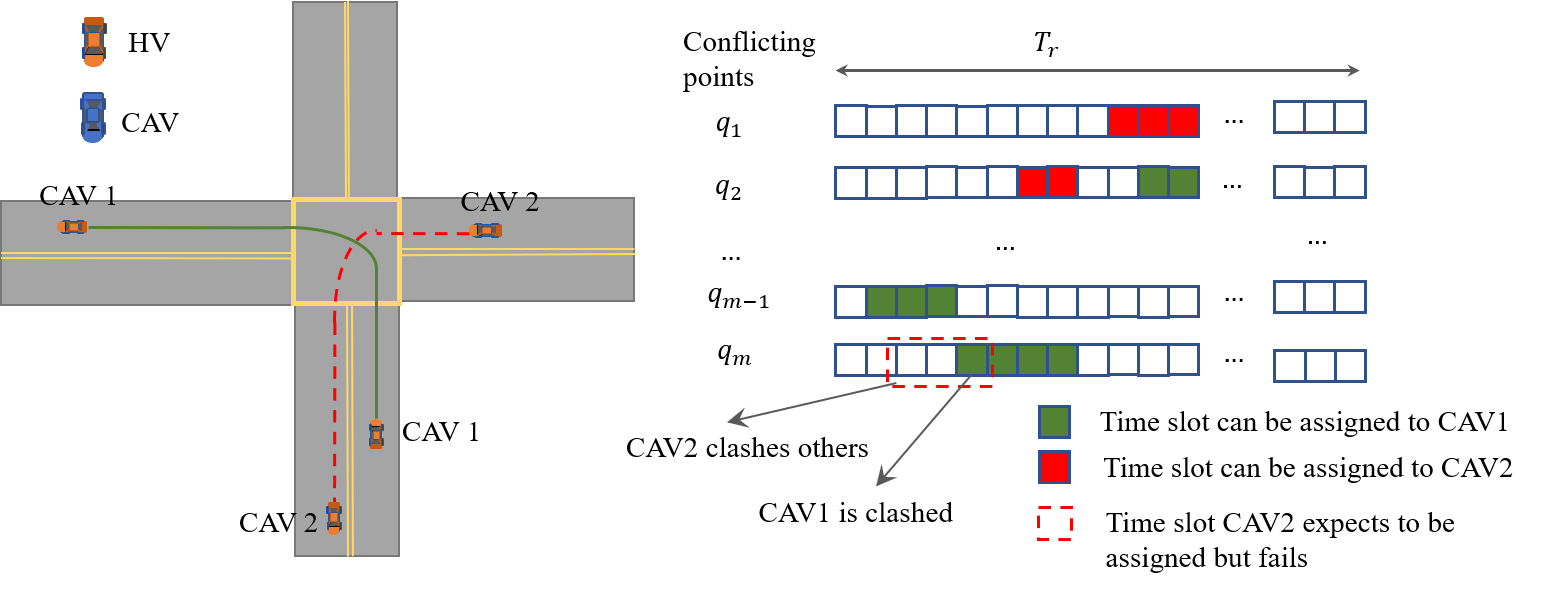}
    \caption{clash and be clashed}
    \label{fig:fig5}
\end{figure}

In principle, for each time block allowed to pass, as long as it has not been assigned to a specific vehicle, then in the order of assignment, vehicles are free to try to request an assignment for themselves, noting that assignment can only be done if all conflict points are not in conflict with other vehicles. When a new assignment cannot be completed at the current time step, vehicles that have already been assigned at the previous time step are asked to maintain their original assignment status, which means the action from deep RL will not be taken for the CAV, while vehicles that have just entered the controlled area are temporarily kept in this "conflicted and unassigned" status until new assignment starts at the next time step. We define two kinds of status: "clash" and "be clashed", where the former represents the ego vehicle requests for a new time slot but be in conflict with the vehicle occupying that slot before, while the latter means other vehicle requests to pass from the time slot the ego vehicle has been assigned, as shown in \ref{fig:fig4} and \ref{fig:fig5}. The order of assignment is conducted according to the distance from the last conflict point of its trajectory, with the closest vehicles being assigned first.

Along the viewpoints above, a hierarchical deep RL framework is proposed, and the agent at upper level push the vehicle who are clashed by other vehicles forward to leave more space for other vehicles to cross, while the agent at lower level prevent causing new conflicts or collision from the accelerated trends at upper level. Further,  when the vehicle is still outside the intersection, it can constantly adjust its speed to prevent the possibility of future conflicts when crossing the intersection and seek to cross the intersection in a shorter period. When the vehicle is in the intersection, its first priority is to avoid collisions. Since the learning tasks and interaction modes are different, we set two groups of reinforcement learning agents, one group adjusting the vehicle's speed outside the intersection and the other group activating after entering the intersection, as shown in \ref{tab:tab1}. When the vehicle located within the outside box and the inside box of \ref{fig:fig3}, agent 1 and agent 2 provide guidance for it, while agent 3 and agent 4 direct its behavior after entering the inner box.

\begin{table}[h]
    \caption{Definition of multiple agents}
    \centering
    \begin{tabular}{ccccc}
    \toprule[1.5pt]
    \makebox[0.15\textwidth][c]{Agent ID} & \makebox[0.15\textwidth][c]{Action Length}& \makebox[0.15\textwidth][c]{Observation} & \makebox[0.15\textwidth][c]{Function} & \makebox[0.15\textwidth][c]{Action span(m/s)} \\ \midrule[1pt]
    1          &   $2s$        & Global coordinates        & Prevent being clashed          & {[} 0 ,  2  {]}    \\
    2          &   $0.5s$        & Global coordinates        & Prevent clashing others  & {[} -0.55 , 0.4 {]}   \\
    3          &   $2s$        & Relative coordinates      & Prevent being collided         & {[} 0 ,  2  {]}  \\
    4          &   $0.5s$        & Relative coordinates      & Prevent colliding others  & {[}-0.55, 0.4 {]}  \\  
    \bottomrule[1.5pt]
    \end{tabular}
    \label{tab:tab1}
\end{table}

As shown in table \ref{tab:tab1}, the first two agent observe the environment globally to find an optimal time to safely cross the intersection and the latter two do a local observation to acquire collision-free actions. Follow this line, the states of environment of first 2 agents are generated globally and that of latter two are in relative coordinates of ego vehicle as shown in \ref{fig:fig6}, and the states of agents are expressed as:

\begin{figure}[t]
    \centering
    \includegraphics[width=\textwidth]{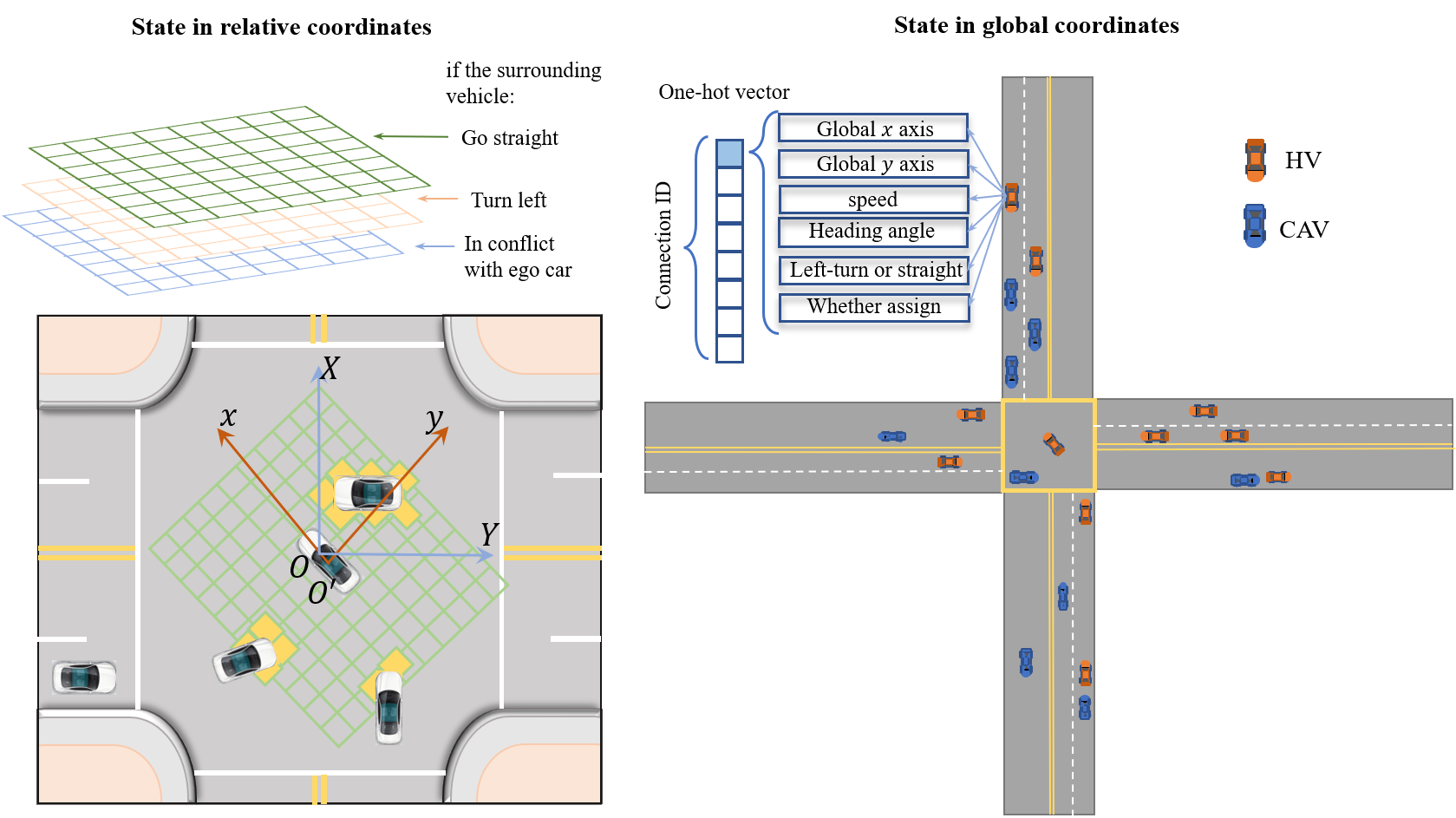}
    \caption{States of reinforcement learning}
    \label{fig:fig6}
\end{figure}

\begin{equation}
    \begin{cases}
      s_{1,t} = [x_t,G_t]\\
      s_{2,t} = [x_t,G_t,a_{1,t}] \\
      s_{3,t} = [x_t,R_t]\\
      s_{4,t} = [x_t,R'_t,a_{3,t}]
    \end{cases}
\end{equation}

where $x_t$ includes the information of ego vehicle itself about global coordinates, absolute speed, heading angle, whether left-turn,etc., then the information is put into a one-hot cell only the corresponding position of the connection ego vehicle drives in is non-zero, and $G_t$ consists of information of other vehicles about global coordinates, absolute speed, heading angle, whether left-turn, and whether in conflict with ego vehicle. Owning to the size of state should be fixed, we take nine vehicles closest to the intersection for each connection $C$. Defining the number of connections at intersection as $N_c$, then the size of external observation of the first two agents is $N_c\times9\times 6$, with $9$ the number of vehicles in each connection and $6$ the number of each vehicle's parameters, and the size of representative vector of ego vehicle $N_c \times 6$. As for $R_t$ and $R'_t$, they represent the occupation of adjacent space by surrounding vehicles with matrices, with size of $L_r\times L_r \times 3$, and $L_r$ means the side length of receptive field. Each element of $R_t$ or $R'_t$ is zero for non-occupation or speed of the vehicle which occupies that grid. As the behaviors of left-turn vehicles are quite different with vehicles in straight direction. We divide the straight and left-turn surrounding vehicles into two matrix, and if the surrounding vehicle is in conflict with ego vehicle, we also put it into the third matrix, which makes up $3$ in the dimension of $R_t$ and $R'_t$. Besides, the agents at lower level desire to know the actions from agents at upper level, thus $a_{1,t}$,$a_{3,t}$ refers to the action selection from agent 1 and agent 3 at this time step. All these observed variables are flatten to vectors and concatenated as state vectors.

The action of each step involves evenly distributed effect of trend from agent at upper level and the action selected by agent at lower level, expressed as:

\begin{equation}
    a_t = 
    \begin{cases}
        a_{t,1}/N + a_{t,2},  &\text{ if CAV in optimization area}\\
        a_{t,3}/N + a_{t,4},  &\text{ if CAV enters intersection}
    \end{cases}
\end{equation}
where $a_{t,i}$ means the action of agent $i$ at time $t$, specifically represent the speed change of ego vehicle at given time-step, and $N$ denotes the multiplier of action length at upper level to the action length at lower level. Considering the vehicle dynamics conditions, $a_t$ should satisfy:
\begin{equation}
    \begin{cases}
        -5 \times \Delta t \leq a_t \leq 2.5 \times \Delta t\\
        v_{min} \leq (a_t + v) \leq v_{max}
    \end{cases}
\end{equation}
where $\Delta t$ is the minimum action length, and $v,v_{min},v_{max}$ is the minimum, current, and maximum velocities of the controlled vehicle. Here the acceleration and emergent deceleration of vehicle is restricted to $2.5m/s^2$ and $-5m/s^2$,respectively. Besides, if adaptive clashing is generated according to \ref{fig:fig4}, $a_t$ will be set to $0$ which means the controlled vehicle will maintain its original speed.

In the hierarchical RL framework, reward at each simulated step has different definition. For the agents at upper level, the step reward is defined as:
\begin{equation}
    r'_t (s_t, r_t) = 
    \begin{cases}
        -1,   &\text{if CAV is still in intersection and none of situation below happens}\\
        -300,  &\text{if CAV is be clashed by other vehicles in expected reservation timetable}\\
        50,   &\text{if CAV leave the intersection}\\
        0,    &\text{after CAV leave the intersection}
    \end{cases}
\end{equation}

The actual reward of agents at upper level is the sum of all step rewards during the feedback period.

\begin{equation}
    r_t(s_T,r_T) = \sum^{N}_{i=1} r'_{t-N+i}
\end{equation}

while for the agents at lower level, the reward is defined as:
\begin{equation}
    r_t (s_t, r_t) = 
    \begin{cases}
        -50,  &\text{if CAV will clash other vehicles in expected reservation timetable}\\
        0,    &\text{otherwise}
    \end{cases}
\end{equation}

From the design of reward, it can be seen the upper level expects to help other vehicles finish assignment in timetable and let ego vehicle pass the intersection as fast as possible. While the lower level expects to avoid conflicting or colliding with others actively. As the collisions rarely happen in the well-trained model, thus we give that state a higher punishment.

\subsection{Human Driver}
The behavioral model of human drivers is adopted intelligent drivel model (IDM), which can be expressed as:
\begin{equation}
a_{I D M}^{(k)}(t)=a_{\max }\left[1-\left(\frac{v^{(k)}(t)}{v_{0}^{(k)}}\right)-\left(\frac{s_{k}^{(k, k-1)}(t)\left(v^{(k)}(t), \Delta v^{(k, k-1)}(t)\right)}{s^{(k, k-1)}(t)}\right)\right]
\end{equation}
\begin{equation}
s_{*}^{(k, k-1)}\left(v^{(k)}, \Delta v^{(k, k-1)}\right)=s_{0}+v^{(k-1)} T+\frac{v^{(k-1)} \Delta v^{(k-1)}}{2 \sqrt{a_{\max } b}}    
\end{equation}
Where $a_{max}$ denotes the maximum acceleration of ego vehicle, $v^{(k)}$ denotes the current velocity and $v_{0}^{(k)}$ represents the expected velocity of ego car. $s^{(k,k-1)}$ and $s_*^{(k,k-1)}$ means the current heading and expected heading between ego car and front car. $b$ is for comfortable deceleration.

\section{Experiments}
\subsection{Description of Experiments}
To verify the effectiveness of proposed model, We conduct simulations in the widespread microscopic traffic simulator, SUMO. The test scenarios are configured as 4-way-8-lane unsignalized intersection with straight-left lanes and straight-right lanes, as shown as \ref{fig:fig3}. The parameter settings of the proposed model and the simulated environment are given in table \ref{tab:tab3}. We compare the performance of the proposed model and baselines at low, medium, and high flow rate with flow rate configured to $450 veh/lane/h$, $900 veh/lane/h$ and $1200 veh/lane/h$,respectively. And the penetration rate of human driving vehicles is set to 0.8.

\begin{table}[h]
    \caption{Details of Parameters}
    \centering
    \begin{tabular}{ccl}
    \toprule[1.5pt]
    \makebox[0.15\textwidth][c]{Parameters}    & \makebox[0.15\textwidth][c]{Values}     & \makebox[0.3\textwidth][c]{Explanations}                     \\ 
    \midrule[1.0pt]
    $l_v$         & $5.0m$     & The length of vehicles                               \\
    $w_v$         & $1.8m$     & The width of vehicles                                \\
    $a^{+}_{max}$ & $2.5m/s^2$   & The maximum acceleration of vehicles               \\
    $a^{-}_{max}$ & $-5m/s^2$   & The maximum deceleration of vehicles                \\
    $v_{max}$     & $15m/s$    & The maximum speed of vehicles in the scenarios       \\
    $N_{epoch}$   & $100$      & The maximum number of training epochs                \\
    $M_{mem}$     & $500000$   & The size of replaying memories of each SAC           \\
    $t_{action}$  & $0.2s$     & The agent's unit action step                         \\
    $B$           & $256$      & The batch size of data input into neural networks    \\
    $\gamma$      & $0.99$      & The discount factor for reward                      \\
    $ \alpha $    & $0.2$      & Temperature parameter for importance of entropy      \\
    $ lr $        & $0.0003$   & Learning rate of the neural networks                 \\ 
    \bottomrule[1.5pt]
    \end{tabular}
    \label{tab:tab3}
\end{table}

\subsection{Performance Evaluation}
We select the metrics of experiments from aspects of safety, efficiency, and fuel consumption:

(1) $N_{col} := $ the number of collisions per hour.

(2) $t_{cross} :=$ the average crossing time from vehicles entering intersection area to leaving the intersection. The boundary of intersection area is defined $30m$ away from the stop line.

(3) $std_t :=$ the standard deviation of $t_{cross}$. The metric reflects the fairness of right of road for different vehicles.

(4) $F :=$ the average fuel consumption per vehicle.

\subsection{Method Comparison}
Six methods are employed for comparison with the proposed model.

(1) Fixed signal timing control: pre-defined signal phases with fixed order and duration change periodically in a round-robin manner. The cycle of signal in this experiment is configured to 120s.

(2) Longest-Queue-First (LQF) timing signal: As a robust method introduced in literature review, LQF always permits the vehicles in the directions with longest queue to cross intersection first.

(3) FCFS with virtual traffic light (VTL): The CAvs follows the principle of FCFS, and HVs follow its front CAV to cross the intersection.

(4) FCFS with virtual traffic light (platoon): On the basis of FCFS with VTL, platoons are formed based on the proximity of vehicles' crossing time to enhance the efficiency, the maximum size of platoon is constant and defined as 8 in our experiments.

(5) Distributed SAC without hierarchical adversarial mechanism: As a comparison of the proposed method, the SAC model without hierarchical adversarial mechanism is employed to guide each CAV to cross the intersection.

\subsection{Results and Analysis}
The table \ref{tab:tab4}-\ref{tab:tab6} give the performance of the proposed model and baselines. From these tables, it is seen that the hierarchical adversarial mechanism can significantly improve the safety of distributed reinforcement learning in the AIM problem, by reducing the collision number from several hundred to a lower level. And the HARL also outperforms other baselines in the passage efficiency. As seen in the graphical interface of the simulation, the CAVs make effort to leave appropriate gaps for other vehicles to cross the intersection and weave through the gaps in as orderly a manner as possible, making full use of the space-time available at the intersection compared to the rule-based method like traffic signals and FCFS. The LQF keeps permitting the request from the longest queue, making it gain a relative reasonable crossing order, but the granularity of the definition of conflicts is the whole intersection, which wastes opportunities that would have allowed some vehicles in conflicting directions to weave in and out through gaps in traffic flows. Thus the proposed model achieve 33.38\%- 55.04\% improvement compared with the LQF method. The method of RL without HARL framework cannot figure out a proper chance to cross for each CAV, and does not slow corresponding CAV down which gives a a relative shorter passage time in some cases, but the collisions threat the safety of passengers which would not be allowed in the real cases.

\begin{table}[h]
    \caption{The results under low traffic flow}
    \begin{tabular}{l|cccc}
    \toprule[1.5pt]
                      & \makebox[0.15\textwidth][c]{$C$}  & \makebox[0.15\textwidth][c]{$t_{cross}(s)$}     & \makebox[0.15\textwidth][c]{$std_t(s)$}   & \makebox[0.15\textwidth][c]{$F(ml/veh)$}     \\ 
                      \midrule[1.0pt]
    Fixed time signal & 0  & 46.97 & 48.24 & 59.34 \\
    LQF signal        & 0  & 13.80  & 9.64  & 19.81 \\
    FCFS-VTL          & 0  & 43.86 & 45.0 & 52.93 \\
    FCFS-VTL(Platoon) & 0  & 48.21 & 44.05 & 58.35 \\
    single-agent RL    & 185 & 8.10 & 0.90  & 8.02 \\
    Proposed method   & 0  & 6.57    & 1.89     & 11.11    \\ 
    \bottomrule[1.5pt]
    \end{tabular}
    \label{tab:tab4}
\end{table}

\begin{table}[h]
    \caption{The results under medium traffic flow}
    \begin{tabular}{l|cccc}
    \toprule[1.5pt]
                      & \makebox[0.15\textwidth][c]{$C$}  & \makebox[0.15\textwidth][c]{$t_{cross}(s)$}     & \makebox[0.15\textwidth][c]{$std_t(s)$}   & \makebox[0.15\textwidth][c]{$F(ml/veh)$}     \\ 
                      \midrule[1.0pt]
    Fixed time signal & 0  & 35.44 & 48.80 & 45.13 \\
    LQF signal        & 0  & 15.67  & 8.96  & 22.14 \\
    FCFS-VTL          & 0  & 42.18 & 38.97 & 51.94 \\
    FCFS-VTL(Platoon) & 0  & 54.28 & 64.72 & 80.00 \\
    single-agent RL    & 688 & 8.34 & 1.39  & 9.11 \\
    Proposed method   & 0  & 10.44   & 15.15    & 21.67    \\ 
    \bottomrule[1.5pt]
    \end{tabular}
    \label{tab:tab5}
\end{table}

\begin{table}[h]
    \caption{The results under high traffic flow}
    \begin{tabular}{l|cccc}
    \toprule[1.5pt]
                      & \makebox[0.15\textwidth][c]{$C$}  & \makebox[0.15\textwidth][c]{$t_{cross}(s)$}     & \makebox[0.15\textwidth][c]{$std_t(s)$}   & \makebox[0.15\textwidth][c]{$F(ml/veh)$}     \\ 
                      \midrule[1.0pt]
    Fixed time signal & 0  & 33.79 & 49.99 & 43.43 \\
    LQF signal        & 0  & 19.33  & 10.24  & 26.10 \\
    FCFS-VTL          & 0  & 45.82 & 49.48 & 57.42 \\
    FCFS-VTL(Platoon) & 0  & 51.10 & 69.08 & 67.74 \\
    single-agent RL    & 996 & 8.01 & 0.93  & 8.37 \\
    Proposed method   & 11  & 8.69   & 7.28     & 16.23    \\ 
    \bottomrule[1.5pt]
    \end{tabular}
    \label{tab:tab6}
\end{table}

\begin{figure}[t]
    \centering
    \includegraphics[width=\textwidth]{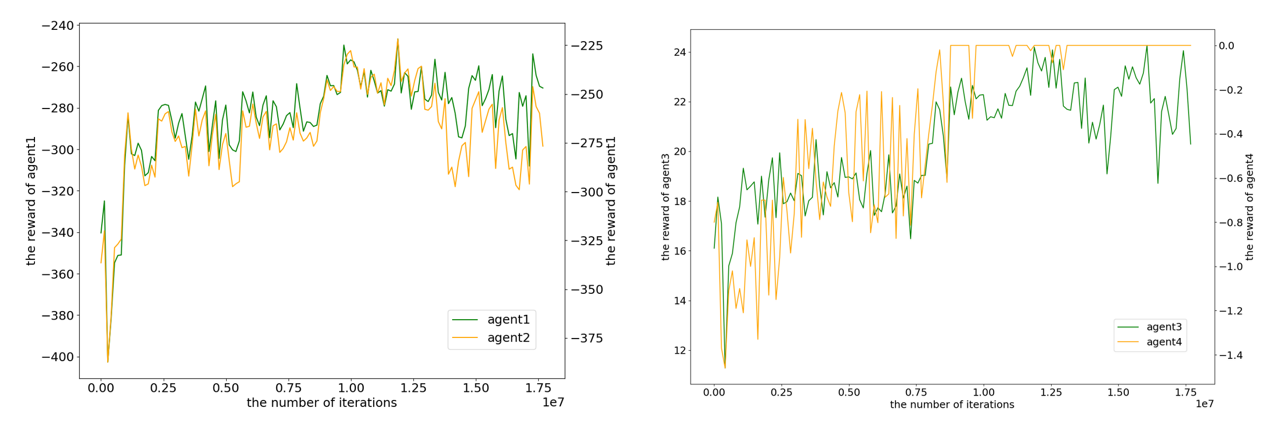}
    \caption{The reward curve of agent 1 and agent 2}
    \label{fig:fig7}
\end{figure}

Figure \ref{fig:fig7} shows the training process of the proposed model, it shows the average reward of all CAVs crossing the intersection per 10000 iterations. It is seen that the reward of the proposed model enhance significantly. The rewards of the different agents show a high positive correlation, and we hypothesize that the reverse reward caused by adversarial action are erased in the average returns over period of time, and the average reward across period are mainly related to the fluctuations in traffic flows and the varying interaction patterns generated by multiple arriving participants at intersection.

\section{Conclusion}
This paper develops a novel hierarchical adversarial framework for reinforcement learning in AIM. The agent in the upper level guide CAV to avoid being clashed and the agent in the lower level restrict the change of speed so that it will not cause a new conflicts with other vehicles. Agents are categorized into two groups with the first group guide each CAV to adjust their time to passage before entering the crossing zone of intersection, the other group takes the duty for safety of CAV at intersection when interweaving. The main contribution of this paper is to design a adversarial structure which reaches a balance between moving forward and yielding, which is one of the major reasons cause collisions in the distributed RL. By differentiating the period of observation, action and reward, the proposed model find the Pareto optimum between avoiding being collided and causing new collisions, then a satisfactory scheduling scheme at intersection is formed.

In our experiment, we compare the proposed model and several widely utilized base structure in AIM, and prove the outperforming performance in efficiency. Then the same Soft Critic-Actor model with and without HARL framework compared in the experiment, and it is seen the framework significantly improve the safety at AIm with the framework. The further work can be extended to a larger coordination in AIM with multiple intersections or more highly-interactive MARL scenarios in other domains.

%Bibliography
\bibliographystyle{unsrt}  
\bibliography{references}

\end{document}